\documentclass[aps,preprint,eqsecnum]{revtex4}

\usepackage{graphicx}

\usepackage{epsf}

\newcommand{\re}[1]{(\ref{#1})}

\newcommand{\up}{\uparrow}

\newcommand{\dn}{\downarrow}

\newcommand {\dis}{\displaystyle}

\newcommand{\beg}{\begin{equation}}
\newcommand{\en}{\end{equation}}

\newcommand{\eps}{\epsilon}

\begin{document}

\title{Integrable dynamics of coupled Fermi--Bose condensates}

\author{Emil A. Yuzbashyan$^{1}$ }

\author{Vadim B. Kuznetsov$^2$}


\author{Boris L. Altshuler$^{3}$}

\affiliation{ \phantom{a} \vspace{0.1cm}
\centerline{$^1$Center for Materials Theory, Department of Physics and Astronomy,}
\centerline{ Rutgers University,
Piscataway, New Jersey 08854, USA}
\centerline{$^2$Department of
Applied Mathematics, University of Leeds, Leeds, LS2 9JT, UK}
\centerline{$^3$Physics
Department, Princeton University, Princeton, NJ 08544, USA} }




\begin{abstract}

We study the mean-field dynamics of a fermionic condensate interacting with a
single bosonic mode (a generalized Dicke model). This problem is integrable and can be mapped onto a corresponding BCS problem. We derive the general solution
and a full set of integrals of motion for the time evolution of coupled Fermi-Bose condensates. The present paper complements our earlier study of the dynamics of
the BCS model (cond-mat/0407501 and cond-mat/0505493).
Here we provide a self-contained introduction to the variable separation method, which enables a complete analytical
description of the evolution of the generalized Dicke, BCS, and other similar
models.
\end{abstract}

\maketitle

\section{Introduction}

 A number of closely related non-stationary  problems have come up recently in different contexts.  In these problems the goal is to describe the dynamics of a  many-body system following a sudden perturbation that drove the system out of an equilibrium. The system in question can be a BCS
superconductor\cite{ander1,galaiko,volkov,galperin,shumeiko,barankov1,simons,amin,BCS,shortBCS,leggett},
 coupled Fermi-Bose condensates\cite{andreev,barankov}, or a single electronic spin  interacting with many nuclear spins (the central spin model)\cite{stamp,khaetskii,schliemann,nazarov1,coish,koka,lukin}. A common feature of all these problems is that they  can be formulated in terms of spin Hamiltonians, which  belong to a class of  integrable  systems known as Gaudin magnets\cite{gaudin,sklyanin,vadim}. It turns out that this fact enables one to solve for the evolution of all these systems using the same approach.

In the present paper we focus on a particular example of such a problem. Namely, we study the dynamics of a fermionic condensate interacting with a single bosonic mode. Recent studies of this problem\cite{andreev,barankov} were motivated by experiments\cite{regal,zwierlein1,kinast,zwierlein2} on cold fermion pairing. The idea was that near the Feshbach resonance \cite{feshbach1,feshbach2} these systems can be modeled by a fermionic condensate of atoms strongly coupled to the lowest energy bosonic mode.

 We assume that the system has been prepared in a nonequilibrium state at $t\le0$ and study the subsequent evolution for $t>0$. Our main result is a complete solution for the dynamics of the system. We also  derive  a full set of its integrals of motion. It turns out that the mean-field time evolution of the coupled Fermi-Bose
condensates can be mapped  onto the corresponding BCS problem, which was solved previously in Refs.~\onlinecite{BCS,shortBCS}.
For this reason,
the emphasis in the present paper is on this mapping and detailed or alternative derivations that have been largely omitted in
Refs.~\onlinecite{BCS,shortBCS}.

The fermion-boson condensate is described by the following Hamiltonian:
\beg
\hat H=\sum_{j,\sigma}\eps_{j} \hat c_{j \sigma }^\dagger \hat
c_{j \sigma}+\omega \hat b^\dagger \hat b +g\sum_j \left( \hat
b^\dagger \hat c_{ j\dn} \hat c_{ j\up}+ \hat b \hat
c_{j\up}^\dagger \hat c_{j\dn}^\dagger\right),
\label{atommol}
\en
where $\eps_j$ are the single-particle energy levels and the operators $\hat c_{j \sigma }^\dagger$
($\hat c_{j \sigma}$)
create  (annihilate)  a fermion of one of the two species $\sigma=\up$ or $\dn$
 in an orbital eigenstate of energy $\eps_j$. Eigenstates
 $|j\up\rangle$ and $|j\dn\rangle$ are related by the time-reversal symmetry\cite{ander}. For example, if the
 single-particle potential
 is translationally  invariant, $|j\up\rangle=|{\bf p}\up\rangle$ and $|j\dn\rangle=|-{\bf p}\dn\rangle$.
Operators $\hat b^\dagger$ ($\hat b$) create (annihilate) quanta of the bosonic field.

We study the dynamics of the fermion-oscillator model \re{atommol} in the mean-field approximation. This amounts to treating the bosonic field
classically, i.e. replacing operators $\hat b^\dagger$ and $\hat b$ with $c$-numbers in the Heisenberg equations of
motion for Hamiltonian \re{atommol}. This procedure is expected
to be exact as long as the bosonic mode is macroscopically populated.   It turns out that in this approximation the dynamics coincides with that of a classical Hamiltonian system, which can be mapped  onto the corresponding BCS problem (see below).  The classical dynamical variables are the time-dependent quantum-mechanical expectation values
$\langle \hat c_{ j\dn} \hat c_{ j\up}\rangle$, $\langle \hat b\rangle$, and
$\langle\sum_\sigma \hat c_{j \sigma }^\dagger \hat
c_{j \sigma}\rangle$.
We will see below that the mean-field approximation is also equivalent to the usual way of obtaining the classical limit -- by
replacing operators with classical variables and their commutators with Poisson brackets.

 Here we assume that the number of energy levels, $n$, interacting with the bosonic field in the system \re{atommol} is arbitrary large, but {\it finite}. In this case the typical evolution
 is {\it quasi}-periodic with $n+1$ incommensurate frequencies. The system exhibits irregular multi-frequency oscillations ergodically exploring the part of
the phase-space allowed by the conservation laws, returning  arbitrarily close to its initial state at irregular time
intervals. For finite $n$, the solution, albeit explicit, is rather complicated. However, it considerably
simplifies in
the thermodynamic limit $n\to\infty$.
In this limit, the return time diverges for most physical initial conditions, while the solution {\it decays} to
a simple limiting dynamics at large times\cite{unpub}. For example, following
an initial decay, $|\langle \hat b(t)\rangle|$ asymptotes either to a constant value
(cf. Refs.~\onlinecite{galaiko,volkov,simons,leggett}) or to an oscillatory behavior characterized by only a few frequencies (cf. Fig.~2 in Refs.~\onlinecite{galperin,barankov1}).

  The  fermion-oscillator model \re{atommol} can be viewed as a {\it generalization} of the Dicke (Tavis-Cummings) model of Quantum Optics. The latter can be
obtained from
\re{atommol} in the zero fermionic bandwidth limit, i.e. when all single-particle levels $\eps_j$ are degenerate,
$\eps_j=\mu$.
To see this, it is useful
to reformulate the model \re{atommol} as a spin-oscillator model using Anderson's pseudospin representation\cite{ander1}. Pseudospins are defined as
\beg
2\hat K_j^z=\hat n_j-1, \qquad \hat K_j^-=\hat c_{j\dn} \hat c_{j\up},\qquad \hat K_j^+=\hat c_{j\up}^\dagger \hat c_{j\dn}^\dagger,
\label{ander}
\en
where $\hat n_j=\sum_\sigma \hat c_{j \sigma }^\dagger \hat
c_{j \sigma}$.
Operators $\hat{\bf K}_j$ have all properties of spin-1/2
on the subspace of unoccupied and doubly occupied (unblocked) levels $\eps_j$.
Singly occupied (blocked) levels do not interact with the bosonic field and are
decoupled from the
dynamics. In terms of pseudospins the Hamiltonian \re{atommol} for $n$ unblocked levels takes the form
\beg
\hat H=\sum_{j=0}^{n-1} 2\eps_j \hat K_j^z+\omega \hat
b^\dagger \hat b+ g\sum_{j=0}^{n-1} \left( \hat b^\dagger\hat K_j^- + \hat
b\hat K_j^+\right).
\label{dicke}
\en

In  the zero bandwidth limit the Hamiltonian \re{dicke} describes an interaction of a  single collective
spin $\hat {\bf T}=\sum_j \hat {\bf K}_j$ with a harmonic oscillator
\beg
\hat H_{Dicke}=2\mu\hat T_z+\omega \hat b^\dagger \hat b+ g(\hat b^\dagger \hat T_-+\hat b \hat T_+).
\label{TC}
\en
This model was suggested by R. H. Dicke in 1953~\cite{dicke}. Its spectrum was obtained
exactly by M. Tavis and F. W. Cummings in 1967~\cite{TC}. In the mean-field approximation it becomes a
classical model for the dynamical variables $\langle\hat{\bf T}(t)\rangle$ and $\langle\hat b(t)\rangle$
 with only two degrees of freedom (see also below). The mean-field solution for the time evolution was outlined by Dicke\cite{dicke} and derived in
 detail by R. Bonifacio and G. Preparata in 1969~\cite{BP}.
 In fact, the resulting classical problem is that of a spherical pendulum.
 The solution for $\langle\hat{\bf T}(t)\rangle$ and $\langle\hat b(t)\rangle$
  is in terms of  elliptic functions and (apart from the azimuthal motion of the pendulum) is simply
  periodic \cite{BP}.

 Remarkably, a more general many-body model (\ref{atommol},\ref{dicke}) also turns out to be integrable.
This was established by M. Gaudin in 1972\cite{gaudin} and later by B. Jurco\cite{jurco}
who used a
different approach. First explicit nonlinear solutions for the mean-field dynamics of the model (\ref{atommol},\ref{dicke})
were constructed in Refs.~\onlinecite{andreev,barankov}. Interestingly, the initial conditions for which these solutions occur are such that the dynamics
of the model (\ref{atommol},\ref{dicke}) reduces to that of the Dicke model \re{TC}, i.e. it can be described in terms of a single {\it collective} spin coupled to the bosonic field. Similar  solutions for
the mean-field dynamics of the BCS model were discovered by R. A. Barankov et. al. in Ref.~\onlinecite{barankov1} and in an unpublished work by V. S. Shumeiko \cite{shumeiko}.

Here we solve for the  dynamics  of the model (\ref{atommol},\ref{dicke}) in the mean-field
approximation for arbitrary initial conditions. We also derive a full set of conservation laws for the
mean-field dynamics.
 The  typical evolution of the system is quasi-periodic with $n+1$ incommensurate  frequencies.
However, for certain special choices of initial
conditions the dynamics is characterized by $m+1<n+1$ incommensurate  frequencies and can be described in terms of $m$ {\it collective} spins coupled to the bosonic field.

Our approach to the mean-field dynamics of the fermion-oscillator model \re{atommol} employs the method
 of {\it separation of variables}. This method was suggested by I. V. Komarov\cite{komarov} and later developed
by E. Sklyanin and V. Kuznetsov\cite{sklyanin,vadim}. It allows us to derive a full set of integrals of motion for both the quantum model
 (\ref{atommol},\ref{dicke}) and its classical counterpart. We also use it to derive and analyze  equations of motion in terms of  separation variables.

 Upon the replacement of the quantum bosonic field $\hat b$ with its time-dependent expectation value, the Hamiltonian
 \re{atommol} becomes similar to the mean-field BCS Hamiltonian. In this analogy, $g\langle \hat b(t)\rangle$
  plays
 the role of the BCS gap $\Delta(t)$. The difference is that in the case of the BCS model $\Delta(t)$ is not an independent dynamical variable, but is related
 to $\langle \hat c_{ j\dn}(t) \hat c_{ j\up}(t)\rangle=\langle \hat K_j^-(t)\rangle$ by the self-consistency
 condition
 $\Delta(t)=g\sum_j\langle \hat K_j^-(t)\rangle$.
 Nevertheless, it turns out to be possible to map the mean-field dynamics of the fermion-oscillator model \re{atommol} with $n$  energy levels onto the corresponding
  BCS problem with $n+1$ levels. This mapping is facilitated by the variable separation and enables us to obtain the general solution for the time dependence of the expectation values
 $\langle\hat {\bf K}_j(t)\rangle$ and $\langle \hat b(t)\rangle$ in terms of {\it hyperelliptic}
 functions\cite{theta,mumford} using the known solution for the mean-field dynamics of the BCS model\cite{BCS,shortBCS}.

The rest of the paper is organized as follows. In section~\ref{classical}, we derive the classical Hamiltonian  that governs the mean-field dynamics of the quantum model (\ref{atommol},\ref{dicke}).  We
show that the dynamics of the resulting classical model is integrable and derive a full set of its conservation
laws in section~\ref{integrability}. In  section~\ref{sov}, we perform a transformation to a new set of variables, which facilitates the solution of the equations of motion. The mapping to the corresponding BCS problem and the general solution for the mean-field
dynamics of the model (\ref{atommol},\ref{dicke}) are derived in section~\ref{solution}. Finally, section~\ref{fewspin} is devoted to an interesting class of particular solutions that include mean-field equilibrium
states and special solutions of Refs.~\cite{andreev,barankov,dicke,BP}.

\section{Classical Model}
\label{classical}

Here we derive the classical Hamiltonian model that describes the mean-field dynamics of the quantum model
(\ref{atommol},\ref{dicke}).

 We start with the Heisenberg equations of motion for the spin-oscillator   model \re{dicke}.
 \beg
 \begin{array}{l}
 \dis \dot {\hat {\bf K}}_j=i\left[ \hat
H, \hat {\bf K}_j \right]=\hat {\bf B}_j \times \hat {\bf K}_j
\qquad \hat {\bf B}_j=(2g\hat b_x, 2g\hat b_y, 2\eps_j),\\
\\
\dis \dot{\hat b}=i[\hat H,\hat b]=-i\omega\hat b-i g\hat T_-\qquad \hat
{\bf T}=\sum_{q=0}^{n-1} \hat {\bf K}_q,\\
\end{array}
\label{spins1}
\en
where the operators $\hat b_x$ and $\hat b_y$ are defined by
$$
\hat b=\hat b_x - i\hat b_y\qquad \hat b^\dagger=\hat b_x + i\hat
b_y.
$$
In the regime when the bosonic mode is macroscopically populated, we can replace operators $\hat b(t)$ and
$\hat b^\dagger(t)$ in Eqs.~\re{spins1}
with $c$-numbers: $b(t)=\langle \hat b(t)\rangle$ and
$\bar b(t)=\langle\hat b^\dagger(t)\rangle$, where $\langle\dots\rangle$ stands for the time-dependent
quantum-mechanical expectation value.
 After this replacement, Eqs.~\re{spins1} become linear in operators.
Taking their quantum-mechanical expectation value, we obtain for ${\bf s}_j(t)=\langle \hat {\bf K}_j(t) \rangle$
\beg
 \begin{array}{l}
 \dis \dot {\bf s}_j={\bf B}_j \times  {\bf s}_j
\qquad  {\bf B}_j=(2g b_x, 2g b_y, 2\eps_j),\\
\\
\dis \dot b=-i\omega b-i gJ_- \qquad {\bf J}=\sum_{j=0}^{n-1} {\bf s}_j.\\
\end{array}
\label{clspins}
\en

An important observation is that Eqs.~\re{clspins} are Hamiltonian equations for the following classical model:
\beg
H=\sum_{j=0}^{n-1} 2\eps_j s_j^z+\omega \bar b b+g\sum_{j=0}^{n-1}\left(\bar b s_j^-+bs_j^+\right),
\label{classdicke}
\en
where $s_j^\pm=s_j^x\pm is_j^y$. This Hamiltonian
governs $n$ {\it classical spins} interacting with a classical harmonic oscillator\cite{class}.

Classical spins are
similar to the angular
momentum and have the same Poisson brackets
\beg
\begin{array}{l}
\dis \left\{ s_j^a, s_j^b\right\}=-\varepsilon_{abc} s_j^c,\\
\\
\dis \{b,\bar b\}=i\qquad \{b_x, b_y\}=\frac{1}{2},\\
\end{array}
\label{poisson}
\en
where $a$, $b$, and $c$ stand for the spatial indexes $x$, $y$, and $z$.
All other Poisson brackets between components of ${\bf s}_j$ and $b$ vanish.
Equations \re{poisson} determine the fundamental Poisson brackets in our problem
in a sense that  Poisson brackets between any other pair of dynamical variables
(functions of $b, \bar b$ and components of ${\bf s}_j$) can be obtained from Eqs.~\re{poisson} using the standard
properties of Poisson brackets\cite{landau}. For example, Eqs.~\re{clspins} can be obtained from the Hamiltonian equations of
motion $\dot {\bf s}_j=\{H, {\bf s}_j\}$ and $\dot b=\{H, b\}$ using Eqs.~\re{poisson} and \re{classdicke}.

Equations of motion \re{clspins} conserve the length of each spin, ${\bf s}_j^2=\mbox{const}$. This fact can also be viewed as a property of the brackets \re{poisson} -- since ${\bf s}_j^2$ Poisson-commutes with
all other dynamical variables, its bracket with the Hamiltonian \re{classdicke} also vanishes.
In the mean-field approximation eigenstates of the Hamiltonian \re{dicke} are
product states. In a pure state  of a spin-1/2, there is always an axis, ${\bf n}$, such that the projection of the
spin onto it is $1/2$, i.e. $\hat{\bf K}_j\cdot{\bf n}=1/2$.  In this case,  ${\bf s}_j^2=\langle \hat {\bf K}_j\rangle^2=1/4$.
Therefore, if the system was in a product state at $t=0$
\beg
{\bf s}_j^2=(s_j^z)^2+s_j^+ s_j^-=\frac{1}{4}.
\label{kas}
\en

If a number of orbitals $\eps_j$ in \re{classdicke} are degenerate, the magnitude of their total classical spin $\sum_{\eps_j=\eps} {\bf s}_j$ is conserved by the equations of motion \re{clspins}. In this case, one can replace the corresponding spins ${\bf s}_j$ with a single classical spin of a larger magnitude,
\beg
\sum_{\eps_j=\eps}\eps_j {\bf s}_j=\eps {\bf S}_{\eps},\quad \mbox{where}\quad
 {\bf S}_\eps=\sum_{\eps_j=\eps} {\bf s}_j,
\label{largespin}
\en
in the Hamiltonian \re{classdicke} and sum over nondegenerate orbitals only. Below we will assume whenever
necessary that such a replacement has been made, i.e. that orbitals $\eps_j$ are nondegenerate.

Finally, comparing Hamiltonians \re{dicke} and \re{classdicke} and brackets \re{poisson} to the corresponding quantum commutators, we note that the
mean-field approximation is equivalent to the standard procedure of going from quantum to classical mechanics by
replacing commutators with Poisson brackets, $i [A,B]\to \{A,B\}$ (here $\hbar=1$).

\section{Integrability}
\label{integrability}

Here we demonstrate that the classical model \re{classdicke}, as well as its quantum counterparts \re{dicke} and
\re{atommol}, are integrable and introduce a useful tool to  analyze  and solve for their dynamics.

The Hamiltonian
\re{classdicke} depends on $2(n+1)$ dynamical variables: 2 angles for each of $n$ spins plus the coordinate and
the momentum of the oscillator\cite{class}. Therefore, its phase space is $2(n+1)$-dimensional and the number of
the degrees of freedom (the number of generalized coordinates) is $n+1$. To show that the classical model \re{classdicke}
is integrable, we have to show that it has $n+1$ independent integrals of motion (see e.g.
Refs.~\onlinecite{arnold,landau}).

Consider the following vector-functions of an auxiliary parameter $u$:
$$
\begin{array}{l}
\dis {\bf L}_j(u)=\frac{{\bf s}_j}{u-\eps_j}\quad j=0,\dots,n-1\\
\\
 \dis {\bf L}_n(u)=\frac{1}{g^2}
\left(
\begin{array}{c}
2g b_x\\
2g b_y\\
2u-\omega\\
\end{array}
\right).\\
\end{array}
$$
It follows from the above definitions and
 Eqs.~\re{poisson} that components of  ${\bf L}_k(u)$
 have the following Poisson brackets:
\beg
\left\{ L_k^a(v), L_k^b(w)\right\}=\frac{\varepsilon_{abc}}{v-w}\left( L_k^c(v)-L_k^c(w)\right)\quad k=0,\dots,n
\label{fundament}
\en

Because components of different ${\bf L}_j(u)$ Poisson-commute with each other, their sum
\beg
{\bf L}(u)=
\frac{1}{g^2}
\left(
\begin{array}{c}
2g b_x\\
2g b_y\\
2u-\omega\\
\end{array}
\right)
+\sum_{j=0}^{n-1} \frac{{\bf s}_j}{u-\eps_j}
\label{lax}
\en
also satisfies Eqs.~\re{fundament}. One can check using only Eq.~\re{fundament} that the square of this vector
commutes with itself at any two values, $v$ and $w$, of the auxiliary parameter, i.e.
\beg
\left\{ {\bf L}^2(v), {\bf L}^2(w)\right\}=0.
\label{integgen}
\en
The function ${\bf L}^2(u)$ acts as a generating function for the model \re{classdicke} and
its integrals of motion. Indeed, evaluating ${\bf L}^2(u)$ from Eq.~\re{lax}, we obtain
\beg
 {\bf L}^2(u)=\dis\frac{(2u-\omega)^2}{g^4}+\frac{4H_n}{\omega g^2}+
 \sum_{j=0}^{n-1}\left[
\frac{2H_j}{g^2(u-\eps_j)}\phantom{,}+\frac{{\bf s}_j^2}{(u-\eps_j)^2}\right].
\label{l2}
\en
where
\beg
H_j=(2\eps_j-\omega)s_j^z +g(\bar b s_j^- + bs_j^+)+
g^2\sum_{k\ne j}\frac{{\bf s}_j\cdot {\bf s}_k}{\eps_j-\eps_k},
\label{hj}
\en
\beg
H_n=\omega\left(\bar b b+ J_z\right)\qquad {\bf J}=\sum_{j=0}^{n-1} {\bf s}_j.
\label{hn}
\en

Hamiltonians $H_j$ defined by Eqs.~(\ref{hj},\ref{hn})
 Poisson-commute with each other
$$
\{ H_j, H_k\}=0\qquad\mbox{for all $j$ and $k$}.
$$
This follows directly from the fact that  Eq.~\re{integgen} holds for any $v$ and $w$.
On the other hand, it is straightforward to verify that the spin-oscillator   Hamiltonian \re{classdicke} is
a linear combination of $H_j$
\beg
H=\sum_{j=0}^{n} H_j .
\label{dickeh}
\en
Since   $H_j$ Poisson-commute with each other, the Hamiltonian \re{classdicke} commutes with all $n+1$
Hamiltonians $H_j$.
Therefore, $H_j$ given by  Eqs.~(\ref{hj},\ref{hn}) are integrals of motion for the classical spin-oscillator
   model, i.e.
this model is integrable.

The above construction based on Eqs.~\re{fundament} and \re{lax} constitutes the so-called Lax vector
representation for the spin-oscillator   model \re{classdicke}. Its main advantages are that it provides a
powerful tool to study the dynamics and can also be extended to quantum models.
In particular, note that integrals \re{hj} and \re{hn} can be trivially generalized to the quantum case
$$
\begin{array}{l}
\dis \hat H_j=(2\eps_j-\omega)\hat K_j^z+g(\hat b^\dagger \hat K_j^-+\hat b\hat K_j^+)+
g^2\sum_{j\ne k}\frac{\hat{\bf K}_j \hat{\bf K}_k}{\eps_j-\eps_k},\\
\\
\dis \hat H_n=\omega\left(\hat b^\dagger \hat b+\hat T_z\right)\qquad \hat {\bf T}=\sum_{q=0}^{n-1} \hat {\bf K}_q .\\
\end{array}
$$
Operators $\hat H_j$ pairwise commute with each other and the spin-oscillator
  Hamiltonian \re{dicke}, which is their linear
combination as in Eq.~\re{dickeh}.

The same construction with classical dynamics in the mean-field approximation and Lax vector for
resulting classical models applies to a number of other models. In particular, a closely related model
is the BCS model
\beg
\hat H_{BCS}={\sum_{j,\sigma}\eps_{j} \hat c_{j \sigma }^\dagger
\hat c_{j \sigma}-g\sum_{j, q} \hat c_{j\up}^\dagger \hat c_{j\dn}^\dagger \hat c_{ q\dn}
\hat c_{ q\up}}.
\label{bcs1}
\en
Notations here are the same as in Eq.~\re{atommol}. In terms of Anderson's spins (see above Eq.~\re{dicke}), the
Hamiltonian reads
\beg
\hat H_{BCS}=\sum_{j=0}^{n} 2\eps_j \hat K_j^z-g\sum_{j,q} \hat K_j^+\hat K_q^-.
\label{bcs2}
\en
The usual BCS mean-field is equivalent to the  procedure (Eqs.~\re{spins1} and \re{clspins})
we  used to derive the classical   Hamiltonian \re{classdicke}, only now the role of $\hat b$ is played
by $\sum_j \hat K_j^-$ (see e.g. Refs.~\onlinecite{barankov1,BCS}). The
mean-field dynamics is described by the following classical Hamiltonian:
\beg
H_{BCS}=\sum_{j=0}^{n} 2\eps_j s_j^z-g\sum_{j,q} s_j^+s_q^-
\label{bcs}
\en
with the Lax vector
\beg
{\bf L}_{BCS}(u)=- \frac{\hat {\bf z}}{g}+\sum_j \frac{{\bf s}_j}{u-\eps_j},
\label{bcslax}
\en
where $\hat {\bf z}$ is a unit vector along the $z$-axis. The Lax vector ${\bf L}_{BCS}(u)$ has the same properties \re{fundament} and
\re{integgen} as before and generates integrals of motion for the BCS model.

\section{Separation of variables}
\label{sov}

 In this section we perform a transformation to a new set of variables, which facilitates the solution of the equations
 of motion~\re{clspins}. The
choice of new variables  naturally follows from the Lax vector construction of the previous section. The new variables
are canonical and also separate the Hamilton-Jacobi equations for the classical Hamiltonian \re{classdicke} in the usual
sense\cite{arnold,landau}.

We define $n$  variables $u_j$ as zeros of
\beg
L_-(u)=L_x(u)-iL_y(u)=\frac{2b}{g}+\sum_{j=0}^{n-1}\frac{s_j^-}{u-\eps_j}.
\label{l-}
\en
Note that the nominator of this expression is a polynomial of degree $n$ and therefore there are $n$ roots
$u_j$ with $j=0,\dots,n-1$. The coefficients of this polynomial are functions of the dynamical variables
$s_j^-$ and $b$. Accordingly, its roots are also functions of $s_j^-$ and $b$ and therefore define a new set of dynamical variables.

Variables $u_j$ play a role of canonical coordinates for the classical Hamiltonian \re{classdicke}. The corresponding
 momenta are defined as $v_j=L_z(u_j)$. Thus, we have
\beg
L_-(u_j)=0,\qquad v_j=L_z(u_j),\qquad j=0,\dots,n-1.
\label{zeros}
\en
Because our system has $2(n+1)$ degrees of freedom, we need two additional variables $u_n$ and $v_n$,
which can be introduced as
\beg
u_n=b,\qquad v_n=\frac{H_n}{\omega b}=\bar b+\frac{J_z}{b},
\label{unvn}
\en
where $H_n$ is given by Eq.~\re{hn}.

Separation variables $(u_j, v_j)$ are canonical, i.e.
\beg
\{u_j, u_k\}=0\qquad \{v_j, v_k\}=0\quad \{v_j, u_k\}=-i\delta_{jk}\quad j,k=0,\dots,n.
\label{canonical}
\en
The first relation in Eq.~\re{canonical} follows from the fact that by Eqs.~\re{zeros} and \re{unvn} variables $u_j$
depend only on mutually Poisson-commuting variables $s_k^-$ and $b$. The second relation follows from
$\{L_z(v), L_z(w)\}=\{L_z(v), b\}=0$. To derive Poisson brackets between $v_j$ and
$u_k$ for $j,k=0,\dots,n-1$; we use the following equation obtained from Eq.~\re{fundament}
\beg
\{L_z(v), L_-(w)\}=\frac{i}{v-w}\left( L_-(v)-L_-(w)\right).
\label{z-}
\en
Evaluating this expression at $v=u_j$ and $w=u_k$, we obtain
\beg
\{L_z(u_j), L_-(w)\}_{w=u_k}=\lim_{w\to u_k}\frac{-i}{u_j-w}L_-(w)=iL'_-(u_k)\delta_{jk}.
\label{in1}
\en
where $L'(u)=\partial L/\partial u$.
On the other hand,
\beg
\{L_z(u_j), L_-(w)\}_{w=u_k}=\{L_z(u_j), L_-(u_k)\}-L'_-(u_k)\{L_z(u_j), u_k\}=-L'_-(u_k)\{L_z(u_j), u_k\},
\label{in2}
\en
where we used $L_-(u_k)=0$. Comparing Eqs.~\re{in1} and \re{in2}, we obtain the last relation in Eq.~\re{canonical}
for $j,k=0,\dots,n-1$.

The original dynamical variables ${\bf s}_j=\langle \hat{\bf K}_j(t)\rangle$ and
$b=\langle \hat b(t)\rangle$ that we are interested in, can be explicitly expressed in terms
of $u_j$ (the inverse map). Indeed, note
that $L_-(u)$ has its zeros at $u=u_j$ and poles at $u=\eps_j$. Therefore, using Eqs.~\re{zeros} and \re{l-},
we can write it as
\beg
L_-(u)=\frac{2b}{g}+\sum_{j=0}^{n-1}\frac{s_j^-}{u-\eps_j}=\frac{2b}{g} \frac{\prod_k(u-u_k)}{\prod_j(u-\eps_j)},
\label{liu}
\en
where we have also used  $L_-(u)=2b/g+O(1/u)$ for large $u$. Equating
 residues  at $u=\eps_j$ and $u=\infty$, we obtain
\beg
s_j^-=\frac{2b}{g} \frac{\prod_k(\eps_j-u_k)}{\prod_{l\ne j}(\eps_j-\eps_l)},
\label{su}
\en
\beg
J_-=\sum_j s_j^-=\frac{2b}{g}\sum_{j=0}^{n-1} (\eps_j-u_j).
\label{j-}
\en

Similarly, using
$$
L_z(u_j)=v_j\qquad L_z(u)=\frac{2u-\omega}{g^2}+O(1/u),
$$
one can express $L_z(u)$ in terms of $u_j$. The $z$-components of classical spins, $s_j^z$, are residues
of $L_z(u)$ at $u=\eps_j$ (see Eq.~\re{lax})
\beg
s_j^z=s_j^-\left[ \frac{\eps_j-\omega/2+\sum_k(u_k-\eps_k) }{bg}+
\sum_k\frac{v_k}{(\eps_j-u_k) L'_-(u_k)} \right].
\label{szu}
\en

It also follows from Eq.~\re{zeros} that  $v_j^2={\bf L}^2(u_j)$. This allows us to express variables $v_j$ through $u_j$ and the integrals of motion $H_j$
\beg
 v_j^2=\dis \frac{(2u_j-\omega)^2}{g^4}+\frac{4H_n}{\omega g^2}+\sum_{k=0}^{n-1}\left[
\frac{2H_k}{g^2(u_j-\eps_k)}\phantom{,}\frac{{\bf s}_k^2}{(u_j-\eps_k)^2}\right] \quad j=0,\dots,n-1
\label{vj}
\en
\beg
v_n=\frac{H_n}{\omega u_n}=\frac{H_n}{\omega b}.
\label{vn}
\en
Thus, to determine the evolution of  ${\bf s}_j(t)=\langle \hat{\bf K}_j(t)\rangle$ and
$b(t)=\langle \hat b(t)\rangle$ we only need to derive and solve the equations of motion for $u_j(t)$.

\section{Equations of motion for separation variables}
\label{motion}

In order to derive the equations of motion for separation variables $u_j$, we first determine the brackets $u_{l,k}=\{H_k, u_l\}$
for  $l,k=0,\dots,n-1$ using Eqs.~\re{vj} and \re{canonical}. This is done by taking Poisson brackets of both sides
of Eq.~\re{vj}  with $u_l$ and solving the resulting system of linear equations for $u_{l,k}$.

As soon as $u_{l,k}$
are found in terms of $u_j$ and $v_j$, we can use the expression \re{dickeh}
for the    Hamiltonian in terms of $H_j$ to determine $\dot u_j$
\beg
\dot u_j=\{H, u_j\}=\sum_k u_{j,k}.
\label{express}
\en

Taking Poisson brackets of both sides of Eqs.~\re{vj} and \re{vn} with $u_l$, we obtain
\beg
 -i g^2 v_j\delta_{jl}=\sum_{k=0}^{n-1}\frac{u_{l,k}}{u_j-\eps_k}\quad j,l=0,\dots n-1.
\label{flows}
\en
In order to determine $u_{n,k}$ and $u_{l,k}$ for $l,k=0,\dots,n-1$, we need to invert the matrix
\beg
A_{jk}=\frac{1}{u_j-\eps_k},
\label{cauchy}
\en
which is called the Cauchy matrix.

This can be done with the help of the following identity coming from partial fraction decomposition of the LHS:
\beg
\frac{\prod_{l=0}^{n-1} (u-u_l)}{(u-u_p)\prod_{l=0}^{n-1} (u-\eps_l)}=\sum_{j=0}^{n-1}\frac{1}{u-\eps_j}
\frac{\prod_{l\ne p} (\eps_j-u_l)}{\prod_{l\ne j}(\eps_j-\eps_l)}.
\label{fract}
\en
To verify this identity note that both sides have the same poles and that the residues at these poles coincide.
Substituting $u=u_q$ in Eq.~\re{fract}, one derives
\beg
\delta_{qp}=\sum_{j=0}^{n-1}\frac{1}{u_q-\eps_j} \frac{\prod_{l=1}^n (u_p-\eps_l)}{\prod_{l\ne j}(\eps_j-\eps_l)}
\prod_{l\ne p}\frac{\eps_j-u_l}{u_p-u_l}.
\label{frac1}
\en
Hence
\beg
\left(A^{-1}\right)_{jk}=\frac{\prod_{l=1}^n (u_k-\eps_l)}{\prod_{l\ne j}(\eps_j-\eps_l)}
\prod_{l\ne k}\frac{\eps_j-u_l}{u_k-u_l}.
\label{cauchyinv}
\en
It follows from Eqs.~\re{flows} and \re{su} that
\beg
 u_{j,k}=-ig^2v_j\frac{\prod_{l=0}^{n-1}(u_j-\eps_l)}{\prod_{l\ne j}(u_j-u_l)}\frac{gs_k^-}{2b(\eps_k-u_j)}.\\
\label{eqs2}
\en
  Finally, noticing that $0=L_-(u_j)=2b/g+\sum_k s_k^-/(u_j-\eps_k)$ and using Eqs.~\re{express} and \re{vj}, we derive
\beg
\begin{array}{l}
\dis \dot u_j=-\frac{2i \sqrt{Q_{2n+2}(u_j)}}{\prod_{m\ne j}(u_j-u_m)} \quad j=0,\dots,n-1\\
\\
 \dis  \dot b=-2i b\left(\frac{\omega}{2}+\sum_{j=0}^{n-1} \eps_j-\sum_{k=0}^{n-1} u_k\right),\\
 \end{array}
\label{dickeev}
\en
where the {\it spectral polynomial} $Q_{2n+2}(u)$ is defined as
\beg
Q_{2n+2}(u)=g^4{\bf L}^2(u)\prod_{j=0}^{n-1}(u-\eps_j)^2.
\label{spectral}
\en
By Eq.~\re{l2}, the coefficients of $Q_{2n+2}(u)$ depend only on the integrals of motion $H_j$ and parameters $\eps_j$ and $g$. The equation of motion for $u_n=b$ is obtained by substituting Eq.~\re{j-} into Eq.~\re{clspins}.

An almost identical derivation of equations of motion
for separation variables can be performed for the mean-field BCS model \re{bcs} using its vector Lax representation
\re{bcslax} leading to the following equations of motion:
\beg
\begin{array}{l}
\dis \dot u_j=-\frac{2i \sqrt{Q^{BCS}_{2n+2}(u_j)}}{\prod_{m\ne j}(u_j-u_m)} \quad j=0,\dots,n-1\\
\\
 \dis  \dot J_-=-2i J_-\left(g J_z+\sum_{j=0}^n \eps_j-\sum_{k=0}^{n-1} u_k\right),\\
 \end{array}
\label{bcsev}
\en
where the spectral polynomial for the BCS model is defined as
\beg
Q^{BCS}_{2n+2}(u)=\frac{1}{g^2} {\bf L}^2_{BCS}(u)\prod_{j=0}^{n}(u-\eps_j)^2.
\label{qbcs}
\en

\section{The solution}
\label{solution}

Here we obtain the general solution for the mean-field dynamics of the fermion-oscillator (spin-oscillator) model
(\ref{atommol},\ref{dicke}) by mapping it onto the corresponding BCS model. The solution therefore can be read off from
the known general solution of the mean-field BCS problem\cite{BCS}.

By comparing equations of motion \re{dickeev} for the spin-oscillator model with those for BCS model \re{bcsev},
we observe that they coincide upon a replacement
\beg
\sum_{j=0}^n\eps_j+gJ_z\to \sum_{j=0}^{n-1}\eps_j+\frac{\omega}{2},\qquad J_-\to b,\qquad Q^{BCS}_{2n+2}(u)\to
Q_{2n+2}(u).
\label{map}
\en

Thus, the mean-field dynamics of the spin-oscillator   model \re{dicke} with $n$ spins and a bosonic field coincides
with that of the BCS model \re{bcs2} with $n+1$ spins. This allows us to immediately write down the solution for the
time-dependent averages ${\bf s}_j(t)=\langle \hat {\bf K}_j(t)\rangle$ and $b(t)=\langle\hat b(t)\rangle$
\beg
 b(t)=\left[J_-\right]_{BCS}, \qquad [{\bf s}_j(t)]_{\mbox{\small Dicke}}=\left[{\bf s}_j(t)\right]_{BCS}.
\label{bt}
\en
The explicit form of $\left[J_-\right]_{BCS}$ and  $[{\bf s}_j(t)]_{BCS}$ in terms of hyperelliptic functions
can be found in Ref.~\onlinecite{BCS} (see Eqs.~(3.22 -- 3.24) of Ref.~\onlinecite{BCS}).

The dynamics of ${\bf s}_j(t)=\langle \hat {\bf K}_j(t)\rangle$ and $b(t)=\langle\hat b(t)\rangle$ is typical
of a classical integrable system with $n+1$ degrees of freedom\cite{arnold} (recall that $n$ is the number of non-degenerate
single particle levels $\eps_j$ in the fermion-oscillator model \re{atommol}). The typical motion is quasi-periodic with
$n+1$ incommensurate periods. For example, a Fourier transform of $b(t)=\langle\hat b(t)\rangle$ shows
$n+1$ basic frequencies. The system uniformly (ergodically) explores the invariant torus -- the $n+1$-dimensional
portion of the $2n+2$-dimensional phase space allowed by the integrals of motion \re{hj} and \re{hn}, returning
arbitrarily close to the initial point at irregular intervals.

We  found that in the thermodynamic limit $n\to\infty$ the solution
simplifies for most physical initial conditions. The return time diverges  in this limit. The dynamics of
$b(t)$ is particularly simple -- it decays, typically as a power-law, to a steady state where $|b(t)|$ is either constant    or
is characterized by a few independent frequencies\cite{unpub}.
 However, the motion of the spin system still contains a
continuum of frequencies. The final steady state of $|b(t)|$ depends on the initial conditions.

\section{Few spin solutions}
\label{fewspin}

The evolution described in the previous section occurs for most initial conditions and is stable against small perturbations
of the Hamiltonian by the KAM theorem even if these perturbations destroy the integrability. However,
there always exists a set of points of measure zero in the phase space, where the motion is characterized by
only a few incommensurate frequencies, while the stability is not guaranteed \cite{arnold}.
Below we consider a family of
such solutions, which we call {\it $m$-spin solutions} with $m<n$. The reason is that in these cases the dynamics of
$n$ spins and the bosonic field in Hamiltonian \re{dicke} degenerates to that of $m<n$ spins and the bosonic field.

Few spin solutions are constructed by choosing integrals of motion $H_j$ (i.e. the initial conditions)
 so that $2(n-m)$ roots of the spectral polynomial $Q_{2n+2}(u)$ defined in Eq.~\re{l2}
 become double degenerate. Suppose $u=E_0$ is a double root\cite{real} of $Q_{2n+2}(u)$ and
note that, since $Q_{2n+2}(E_0)={\bf L}^2(E_0)=0$, setting e.g. $u_{n-1}(t)=E_0$ solves the equation of motion \re{dickeev}
for the variable $u_{n-1}$. Therefore, we can "freeze" one of the separation variables in this root. Then $L_-(E_0)=0$, and it follows from
$L_z^2(u)+L_-(u)L_+(u)={\bf L}^2(u)$ that $L_z(E_0)=0$.
   Thus, one can factor out $(u-E_0)$ from all components of the Lax vector
and show that it is proportional to the Lax vector of the spin-oscillator   model with $n-1$ spins. Similarly, if there are 2 pairs of double
degenerate roots, the number of spins reduces to $n-2$ etc. This procedure is followed in detail in Ref.~\onlinecite{BCS} using a
different method.

First, let us consider the general case when the number of spins effectively reduces from $n$ to $m<n$. We have
\beg
{\bf L}(u)=\left( 1+\sum_j \frac{d_j}{u-\eps_j}\right) {\bf L}_{\bf t}(u),
\label{red}
\en
where $d_j$ are time-independent constants and ${\bf L}_{\bf t}(u)$ is the Lax vector of the $m$-spin problem
\beg
{\bf L}_{\bf t}(u)=\frac{1}{g^2}\left(
\begin{array}{c}
2g b_x\\
2g b_y\\
2u-\omega'\\
\end{array}
\right)+\sum_{k=0}^{m-1}\frac{{\bf t}_k}{u-\mu_k}.
\label{ls}
\en
The $m$-spin system has its own $m$ arbitrary "energy levels" $\mu_k$. Its dynamics is governed by the spin-oscillator
Hamiltonian \re{classdicke} for $m$ spins with new parameters replacing $\eps_j$ and $\omega$.
\beg
H_m=\sum_{k=0}^{m-1} 2\mu_k t_k^z+\omega' \bar b b+g\sum_{k=0}^{m-1}\left(\bar b t_k^-+bt_k^+\right).
\label{degcl}
\en
Matching the residues at poles at $u=\eps_j$ on both sides of Eq.~\re{red}, we express the original spins ${\bf s}_j$ in terms of {\it collective} spins ${\bf t}_k$
\beg
{\bf s}_j=d_j{\bf L}_{\bf t}(\eps_j).
\label{st}
\en
Constants $d_j$ are determined from the above equation using ${\bf s}_j^2=1/4$. We have
\beg
d_j=\frac{|{\bf s}_j| e_j}{\sqrt{ {\bf L}_{\bf t}^2(\eps_j)}}=\frac{ e_j}{2\sqrt{ {\bf L}_{\bf t}^2(\eps_j)}}\quad
e_j=\pm 1.
\label{bj}
\en

Finally we have to match the residues at $u=\mu_k$ and the $u\to\infty$ asymptotic. This leads to the following
$m+1$ equations
\beg
\begin{array}{l}
\dis 1+\sum_{j=0}^{n-1}\frac{d_j}{\mu_k-\eps_j}=0\quad k=0,\dots,m-1\\
\\
\dis \omega=\omega'-2\sum_{j=0}^{n-1}d_j.\\
\end{array}
\label{const1}
\en
Equations \re{const1} constrain the lengths of new spins ${\bf t}_k$, i.e. the coefficients of the spectral
polynomial $Q_{2m+2}(u)=g^4{\bf L}_{\bf t}^2(u)\prod_{k=0}^{m-1}(u-\mu_k)^2$ of the $m$-spin system (see Eq.~\re{l2}). Indeed, using Eq.~\re{bj}, one can cast the constraints Eq~\re{const1} into the form
\beg
\sum_{j=0}^{n-1}\frac{e_j \eps_j^{l-1} }{\sqrt{ Q_{2m+2}(\eps_j)}}=\frac{2}{g^2}\delta_{lm}\quad l=1,\dots,m
\label{const3}
\en
\beg
\omega'=\omega+2\sum_{k=0}^{m-1}\mu_k-\sum_{j=0}^{n-1} \frac{g^2e_j\eps_j^m}{\sqrt{ Q_{2m+2}(\eps_j)} }.
\label{const4}
\en
To obtain $m$-spin solutions explicitly, one has to choose  parameters $\mu_k$, resolve the $m+1$ constraints \re{const3} and \re{const4} for
the lengths of spins ${\bf t}_k$  and frequency $\omega'$, and solve for the dynamics of the $m$-spin Hamiltonian
\re{degcl}. The dynamics can be obtained from the general solution \re{bt} by replacing $n\to m$ and the set of $\{\eps_j\}$
with $\{\mu_k\}$.

Let us illustrate the construction of $m$-spin solutions by considering the cases $m=0$ and $m=1$ in more detail.

{\bf m=0}. The 0-spin solutions correspond to the mean-field eigenstates of quantum Hamiltonians \re{atommol} and \re{dicke}. They are equilibrium states for the classical Hamiltonian \re{classdicke}.

For $m=0$ the dynamics is governed
by the Hamiltonian $H_0=\omega'\bar b b$ with an obvious solution $b=b_0e^{-i(\omega' t+\phi)}$, where $b_0=|b(t)|=\mbox{const}$. Further, using Eqs.~(\ref{ls}), (\ref{st}), and (\ref{bj}), we obtain
\beg
{\bf L}_{\bf t}^2(u)=\frac{1}{g^4}\left( 4g^2b_0^2+(2u-\omega')^2\right),
\label{l0}
\en
\beg
\langle \hat c_{ j\dn} \hat c_{ j\up}\rangle=s_j^-=\frac{e_j g b}{\sqrt{4g^2b_0^2+(2\eps_j-\omega')^2} }, \quad
\langle \hat n_j\rangle-1=2s_j^z=\frac{2e_j(2\eps_j-\omega')}{\sqrt{4g^2b_0^2+(2\eps_j-\omega')^2} }.
\label{-z0}
\en
The frequency $\omega'$ and the gap $\Delta_0=g b_0$ have to satisfy a single constraint \re{const4} that now reads
\beg
\omega'=\omega-\sum_{j=0}^{n-1}\frac{g^2 e_j}{\sqrt{4g^2b_0^2+(2\eps_j-\omega')^2} }.
\label{gapeq}
\en

We see from Eq.~\re{-z0}, that all spins rotate uniformly around the $z$-axis with a frequency $\omega'$. This rotation
can be eliminated, i.e. $\omega'$ can be set to zero, by an appropriate  choice of the chemical potential. Then, Eq.~\re{gapeq} is the analog of the BCS gap equation. The configuration of spins \re{-z0} corresponds to the product BCS
 wavefunction. The latter can be straightforwardly reconstructed from the knowledge of
 ${\bf s}_j=\langle \hat {\bf K}_j\rangle$ as fixing the average of spin-1/2 uniquely fixes its quantum state. The choice
 of signs $e_j=+1$ for all $j$ corresponds to the ground state, while choosing one of the signs to be $-1$ corresponds to
 an excited state -- pair excitation of the fermionic condensate.

There is also an important class of 0-spin solutions that is obtained by setting $b_0=0$. We see from Eq.~\re{-z0} that in this case all spins are along the $z$-axis, $s_j^z=e_j/2=\pm 1/2$. The constraint \re{gapeq} is now irrelevant, because all
$xy$-components vanish. The choice of signs $e_j=\mbox{sign }\eps_j$, where $\eps_j$ are counted from the Fermi level, yields the Fermi ground state. Other choices correspond to the excitations of the Fermi gas. All these states are
stationary with respect to the mean-field dynamics. For a finite number of pairs, they are non-stationary with respect
to the quantum Hamiltonian \re{atommol}. Thus, their initial evolution is entirely governed by quantum corrections
(cf. Refs.~\onlinecite{BP,leggett,shortBCS}).

{\bf m=1}. In this case the dynamics reduces to that of a single collective spin ${\bf t}$ coupled to an oscillator, i.e.
it is governed by the classical counterpart of the Dicke model \re{TC}
\beg
H_1=2\mu t_z+\omega'\bar b b +g(\bar b t_-+b t_+).
\label{TCclass}
\en
Using Eq.~\re{-z0}, one can express  original spins in terms of the collective spin and the bosonic field
\beg
s_j^-=\frac{e_j(\eps_j-\mu) b+ ge_jt_-}{2g\sqrt{Q_4(\eps_j)} }\quad
s_j^z=\frac{e_j(2\eps_j-\omega')(\eps_j-\mu) +g^2 t_z}{2g^2\sqrt{Q_4(\eps_j)} }.
\label{1spin}
\en
To complete the construction of 1-spin solutions, we need to choose a positively defined polynomial $Q_4(u)$ so that it satisfies two constraints \re{const3} \re{const4} and solve for the dynamics of the classical Dicke model \re{TCclass}. In this case there is only one non-stationary separation variable $u_0$ and equations of motion \re{dickeev} take the following form:
\beg
\begin{array}{l}
\dis \dot u_0^2+4Q_4(u_0)=0,\\
\dot b=-ib(\omega+2\mu-2u_0).\\
\end{array}
\label{1ev}
\en
These equations can be solved in terms of elliptic functions.
This is not surprising, since the model \re{TCclass}
is, in fact, equivalent to a spherical pendulum and its solution has already been obtained in Refs.~\cite{dicke,BP}.

The 1-spin solutions were originally obtained in Refs.~\cite{andreev,barankov} where they were used to describe the evolution beginning from a state
infinitesimally close to the normal ground state.

\section{Conclusion}

In this paper we solved for the mean-field dynamics of the fermion-oscillator model \re{atommol}.  In
the mean-field approximation, the problem reduces to a classical Hamiltonian model.
We derived integrals of motion for both the classical \re{classdicke} and the quantum \re{atommol} models and
showed that the dynamics of the fermion-oscillator model \re{atommol} maps onto that of the BCS model. This was used to derive an explicit general solution for the mean-field dynamics of the fermion-oscillator model with an arbitrary
finite number $n$ of degrees of freedom.

The typical dynamics is quasi-periodic with $n$ incommensurate frequencies. The system ergodically explores the part of
the phase-space allowed by integrals of motion, returning arbitrarily close to its initial state at irregular time
intervals. We have also constructed a class of particular, few-spin, solutions, for which the dynamics reduces to
that of $m$ collective spins governed by the same classical model. The case $m=0$ corresponds to the mean-field eigenstates of the fermion-oscillator
Hamiltonian~\re{atommol}.

An interesting  problem is to obtain and describe the thermodynamic limit, $n\to\infty$, of the solution. For most physical initial conditions the dynamics considerably simplifies in this limit\cite{unpub}. In particular,
we found that
$\langle \hat b(t)\rangle$ decays, typically as a power-law, to a steady state where $|\langle \hat b(t)\rangle|$ is either a constant  (cf. Refs.~\onlinecite{galaiko,volkov,simons,leggett})
or oscillates with few independent frequencies (cf. Fig.~2 in Refs.~\onlinecite{galperin,barankov1}).
 The motion of the spin system still contains a
continuum of frequencies. The steady state of $|\langle\hat b(t)\rangle|$ depends on the initial conditions.

It is also interesting
to study  quantum and finite temperature effects. Quantum corrections to the mean-field dynamics can become
important when the bosonic mode is weakly populated \cite{BP}. For example, normal states where
$\langle \hat b^\dagger \hat b\rangle=\langle \hat c_{j\dn} \hat c_{j\up}\rangle=0$ are stationary in the mean-field
approximation \re{clspins}, but are not stationary with respect to quantum dynamics \re{spins1}.


\begin{thebibliography}{99}
\bibitem{class} The coordinate, $x$, and the momentum, $p$ of the oscillator are related to the fields $b$ and
$\bar b$ as $x=\sqrt{2}b_x=(b+\bar b)/\sqrt{2}$ and $p=\sqrt{2}b_y=(\bar b - b)/(\sqrt{2} i)$.  Then, we have the usual coordinate-momentum (canonical) brackets
$\{x,p\}=1$, while the second term in the Hamiltonian \re{classdicke}
reads $\bar b b=p^2/2+x^2/2$.
\bibitem{real} Note that if the root $E_0$ is real, it has to be degenerate, since according to Eq.~\re{spectral} $Q_{2n+2}(u)\ge 0$ for real $u$. In this case ${\bf L}^2(E_0)=0$ implies $L_z(E_0)=L_-(E_0)$ and, since zeros of
$L_-(u)$ are separation variables, one of them {\it has to be} frozen at $E_0$.
\end{thebibliography}
\end{document}